\documentclass{article}
\usepackage{natbib}
\usepackage{color}
\usepackage{hyperref}
\hypersetup{colorlinks=true,citecolor=blue,  linkcolor=blue,urlcolor=black}
\usepackage{graphicx}
\usepackage{amsmath}
\usepackage{amssymb}
\usepackage{mathtools}
\usepackage{authblk}
\usepackage{setspace}
\usepackage{lineno}
\title{CO$_2$ condensation is a serious limit to the deglaciation of Earth-like planets}

\author[1]{Martin Turbet \\ \href{martin.turbet@lmd.jussieu.fr}{(martin.turbet@lmd.jussieu.fr)}}
\author[1]{Francois Forget}
\author[2]{Jeremy Leconte}
\author[3]{Benjamin Charnay}
\author[4]{Gabriel Tobie}

\affil[1]{Laboratoire de M\'et\'eorologie Dynamique, Sorbonne Universit\'es, UPMC Univ Paris 06, CNRS, 4 place Jussieu, 75005 Paris, France.}
\affil[2]{Laboratoire d'astrophysique de Bordeaux, Univ. Bordeaux, CNRS, B18N, all\'ee Geoffroy Saint-Hilaire, 33615 Pessac, France.}
\affil[3]{LESIA, Observatoire de Paris, PSL Research University, CNRS, Sorbonne 
Universit\'es, UPMC Univ. Paris 06, Univ. Paris Diderot, Sorbonne Paris Cit\'e.}
\affil[4]{Laboratoire de Plan\'etologie et G\'eodynamique, UMR-CNRS 6112, University of Nantes, 2 rue de la Houssini\`ere, F-44322 Nantes, France.}
\begin{document}

\maketitle

\begin{abstract}

It is widely believed that the carbonate$-$silicate cycle is the main agent, through volcanism, to trigger deglaciations 
by CO$_2$ greenhouse warming on Earth and on Earth-like planets when they get in a frozen state. 
Here we use a 3D Global Climate Model to simulate the ability of planets initially completely frozen 
to escape from glaciation episodes by accumulating enough gaseous CO$_2$. 
The model includes CO$_2$ condensation and sublimation processes and the water cycle.
We find that planets with Earth-like characteristics (size, mass, obliquity, rotation rate, etc.) orbiting a Sun-like star 
may never be able to escape from a glaciation era, if their orbital distance is greater than $\sim$~1.27~Astronomical Units 
(Flux~$<$~847~W~m$^{-2}$ or 62$\%$ of the Solar constant), because CO$_2$ would condense at the poles 
-- here the cold traps -- forming permanent CO$_2$ ice caps. This limits the amount of CO$_2$ in the atmosphere and 
thus its greenhouse effect.
Furthermore, our results indicate that for (1) high rotation rates (P$_{\text{rot}}~<~$24~h), 
(2) low obliquity (obliquity$~<~$23.5$^\circ$), 
(3) low background gas partial pressures ($<~$1bar), 
and (4) high water ice albedo (H$_2$O albedo$~>~$0.6), 
this critical limit could occur at a significantly lower equivalent distance (or higher insolation).
For each possible configuration, we show that the amount of CO$_2$ that can be trapped 
in the polar caps depends on the efficiency of CO$_2$ ice to flow laterally as 
well as its gravitational stability relative to subsurface water ice. 
We find that a frozen Earth-like planet located at 1.30~AU of a Sun-like star could store
as much as 1.5, 4.5 and 15~bars of dry ice at the poles, for internal heat fluxes of 100, 30 and 10~mW~m$^{-2}$, respectively. 
But these amounts are in fact lower limits.  
For planets with a significant water ice cover, we show that CO$_2$ ice deposits 
should be gravitationally unstable. They get buried beneath the water ice cover 
in geologically short timescales of $\sim$~10$^4$~yrs, mainly controlled by 
the viscosity of water ice. 
CO$_2$ would be permanently sequestered underneath the water 
ice cover, in the form of CO$_2$ liquids, CO$_2$ clathrate hydrates and/or dissolved in subglacial 
water reservoirs (if any). This would considerably 
increase the amount of CO$_2$ trapped and further reduce the probability of deglaciation.

\end{abstract}

\section{Introduction}

There are geological evidences that Earth had to face multiple episodes of global or quasi-global glaciation, 
at the end of the Archean 2.45-2.22~Gya and during the Neoproterozoic era, 710 and 650~Mya 
\citep{Evan:97,Hoff:98,Hoff:02}.

It is widely believed that the carbonate$-$silicate cycle \citep{Walk:81,Kast:93,Kump:00} was the main agent to trigger deglaciations on ancient Earth.
In particular, on a completely frozen planet, also called 'Hard Snowball', the weathering of CO$_2$ is stopped ; 
continuous volcanic outgassing builds up atmospheric 
CO$_2$ that warm up the climate until liquid water is produced in the equatorial regions.

By extension, this mechanism may be crucial to stabilize the climate of Earth-like exoplanets. 
It is even central for the definition of the classical Habitable Zone \citep{Kast:93,Forg:97,Kopp:13}, which 
assumes that planets can build up CO$_2$ atmospheres (as massive as wanted) 
suitable for the stability of surface liquid water.

Planets near the outer edge of the Habitable Zone may suffer, as Earth did, episodes of complete glaciation. 
Recent work by \cite{Meno:15} has even shown that, in the case of planets lacking land vascular plants, temperate climates 
may not be stable, because at high CO$_2$ partial pressure, the increased weathering rate does not allow temperate stable solutions. 
This effect, which is enhanced for planets weakly irradiated by their star, 
may systematically drive Earth-like exoplanets toward episodes of glaciation.

Can Earth-like planets in the Habitable Zone of their star always escape from episodes of glaciation ?
In this work, we use 3D Global Climate Model (GCM) simulations of snowball planets 
to study the ability of the increased CO$_2$ greenhouse effect resulting from volcanic outgassing 
to drive them out of glaciation. As the CO$_2$ outgassed by volcanoes accumulates in the atmosphere, 
the temperature of condensation of CO$_2$ can exceed the surface temperature of the poles. 
This leads to the trapping of extra outgassed CO$_2$, forming permanent CO$_2$ polar ice caps, 
and thus seriously limits the efficiency of the carbonate$-$silicate cycle. 
This possibility has already been suggested in \cite{Pier:05} and \cite{Pier:11areps}, but has never been explored yet. 
We propose in this paper a detailed study of this scenario.

\section{Method}
\label{method}

We use in this paper the 3-Dimensions LMD Generic Global Climate Model to study the deglaciation of Earth-like planets orbiting 
in a circular orbit around a Sun-like star (Sun spectrum and luminosity) in response to increasing amounts of atmospheric CO$_2$ (from 0.01 to 3~bars), 
and for orbital distances ranging from 1.10 to 1.45 Astronomical Units (AU). Detailed information on the model can be found in Appendix~\ref{appen_lmd_model}.

Our simulations were designed to represent completely frozen Earth-like planet characteristics. 
These include the radius (6370~km), the gravity field (9.81~m~s$^{-2}$), 
the obliquity (23.5$^\circ$) and the rotation speed (7.28$\times$10$^{-5}$ rad~s$^{-1}$).
The roles of obliquity, planetary mass and rotation 
rate on the ability of planets to escape glaciation episodes are discussed in sections~\ref{null_obliquity_section}~and~\ref{sensitivity}.
Eventually, most of the simulations were performed for a uniformly flat topography. 
The effect of topography is discussed in section~\ref{topography}.

All the simulations performed in this study were forced initially in a cold and dry state, assuming:
\begin{enumerate}
 \item A uniform and complete ice cover.
 \item Atmospheric and surface temperatures arbitrarily fixed to 230~Kelvins everywhere.
 \item No water vapor, no clouds.
\end{enumerate}

Depending on CO$_2$ partial pressure, obliquity, or parameterization of clouds, the simulations evolve in different 
steady state climate regimes that are discussed in the next section.

\section{Glaciation escape limited by CO$_2$ atmospheric collapse}

\subsection{Reference case}
\label{ref_case}

\begin{table*}
\centering
\begin{tabular}{ll}
   \hline
              &              \\
   Explored parameters & Values \\
              &              \\
   \hline
              &              \\
   CO$_2$ partial pressures (in bar) & [0.01, 0.1, 0.4, 1.0, 1.5, 2.0, 3.0] \\
              &              \\
   i) Stellar flux (in W$~$m$^{-2}$)  & [1130, 1033, 949, 874, 808, 750, 697, 650] \\
   ii) Flux compared to Earth (S$_{\text{eff}}$) & [0.827, 0.756, 0.695, 0.640, 0.592, 0.549, 0.510, 0.476] \\
   iii) Equivalent distance (in AU) & [1.10, 1.15, 1.20, 1.25, 1.30, 1.35, 1.40, 1.45] \\
              &              \\
   Obliquity  & 0, 23.5$^{\circ}$ \\
              &              \\
   CO$_2$ ice clouds  & radiatively active, inactive \\
              &              \\
   \hline

\end{tabular}
\caption{Physical Parameterizations used for GCM calculations.}
\label{param_experiment}
\end{table*}

\begin{figure}
\begin{center}
\includegraphics[width=15cm]{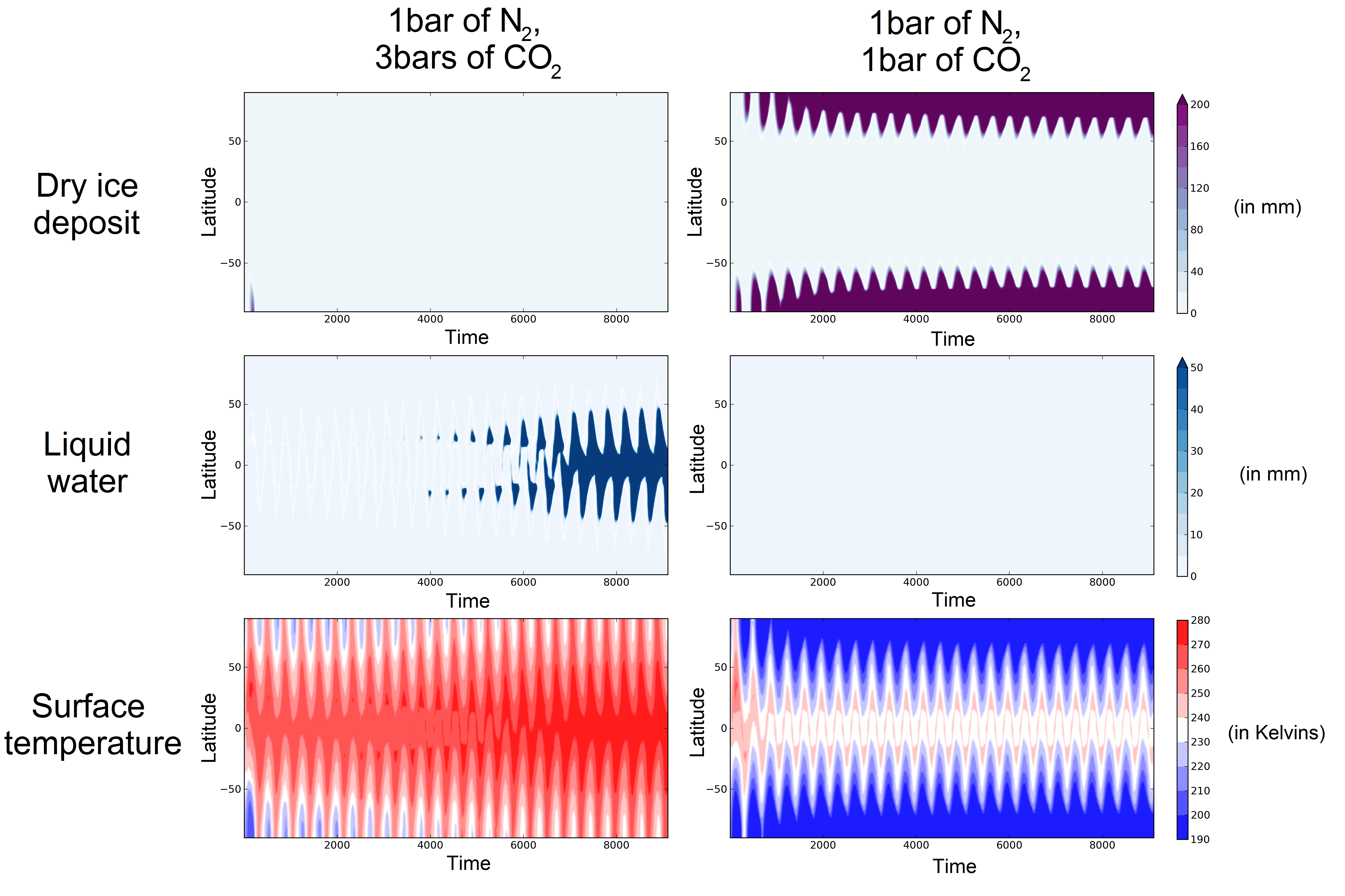}
\caption{From top to bottom: Zonal means of (1) CO$_2$ ice deposit, (2) surface liquid water 
and (3) surface temperatures as a function of time (in Earth days), and for two different initial states. 
On left, the initially cold and dry planet starts with a CO$_2$ partial pressure of 3~bars and 
is able to escape glaciation.
On right, the planet starts with a CO$_2$ partial pressure of 1~bar 
and ends up in atmospheric collapse.
These simulations were run for planets with a 23.5$^{\circ}$ obliquity, radiatively inactive CO$_2$ ice clouds, 
and located at an equivalent orbital distance of 1.30~AU (F~=~808~W~m$^{-2}$).
}
\label{zoology_plots}
\end{center}
\end{figure}

We performed 40 simulations of Earth-like planets (as described in section~\ref{method} -- with a 23.5$^{\circ}$ obliquity 
and radiatively inactive CO$_2$ ice clouds) starting from an initially cold (T~=~230~K everywhere) and global glaciation state 
for multiple irradiation fluxes and CO$_2$ partial pressures (see table~\ref{param_experiment}). 
We fix the N$_2$ partial pressure to be constant and equal to 1~bar, in order 
to be consistent with the case of Earth.
All simulations were run long enough to reach equilibrium\footnote{up to 30 Earth years for the thickest atmospheric configurations.}.

We find that, depending on the CO$_2$ partial pressure and the stellar flux, these simulations can evolve in three different climate regimes:
\begin{enumerate}
 \item The greenhouse effect of CO$_2$\footnote{which is enhanced by the pressure broadening of N$_2$.} is sufficient to raise the surface temperatures in 
equatorial regions above the melting temperature of water ice\footnote{273~K, here.}. 
In this case, the positive feedback due to the decrease of the surface albedo 
(from 0.6 to 0.07) and the greenhouse effect of water vapor drive the planet out of glaciation.
\item The greenhouse effect of CO$_2$ is too weak to trigger a deglaciation. The planet stays in a snowball state.
\item The greenhouse effect of CO$_2$ is too weak to raise the surface temperatures of the poles above the condensation temperature of CO$_2$. 
In this case, atmospheric CO$_2$ collapses and the planet is locked in a global glaciated state, with two permanent CO$_2$ ice polar caps.
\end{enumerate}
Figure~\ref{zoology_plots} shows that, for a given solar flux (808~W~m$^{-2}$ here), the choice of the initial CO$_2$ partial pressure can 
either drive the planet out of glaciation (pCO$_2$~=~3~bars) or cause a permanent collapse of the atmosphere at the poles (pCO$_2$~=~1~bar).

\begin{figure}
\begin{center}
\includegraphics[width=10.5cm]{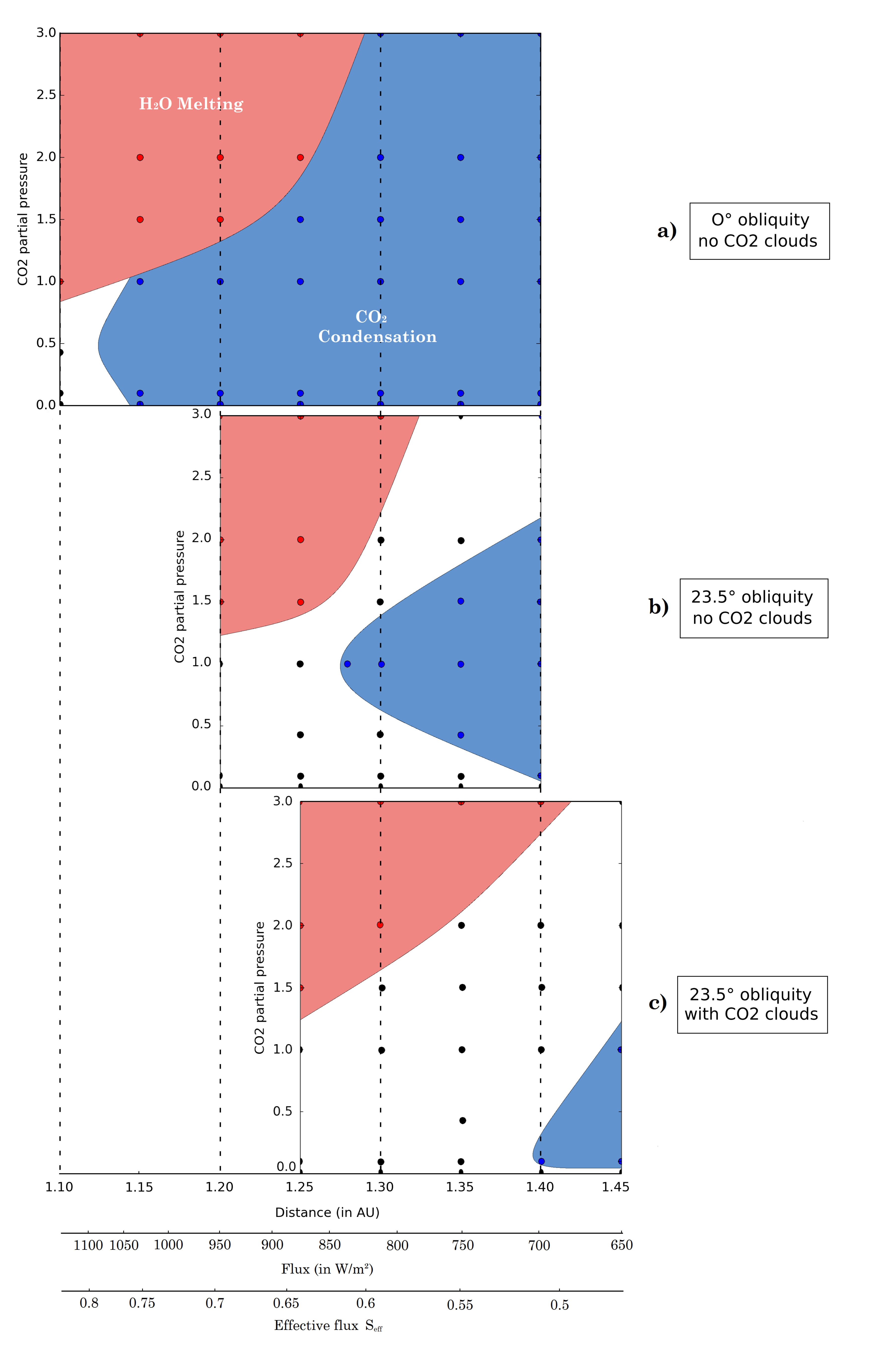}
\caption{Climate regimes reached as function of the equivalent orbital distance from a Sun-like star (in AU) and the CO$_2$ partial pressure, 
assuming a cold start (i.e. snowball state without permanent CO$_2$ ice deposits). 
Diagrams a, b and c were constructed for Earth-like planets with 0$^{\circ}$ obliquity, 23.5$^{\circ}$ obliquity 
(reference simulation) and 23.5$^{\circ}$ obliquity with radiatively active CO$_2$ ice clouds, respectively.
The red color roughly depicts the region where deglaciation is observed. 
The blue region represents glaciated states where CO$_2$ collapses permanently. 
The white region describes cases where none of this two previous conditions were reached.}
\label{condens_plots}
\end{center}
\end{figure}

These results are summarized in Figure~\ref{condens_plots}b, for multiple stellar fluxes and CO$_2$ partial pressures. Each region of the diagram 
denotes a steady state climate regime reached by the planet:
\begin{enumerate}
 \item In red, the planet is partially or totally deglaciated.
 \item In white, the planet is entirely frozen.
 \item In blue, the planet is entirely frozen and gaseous CO$_2$ has permanently collapsed at the poles. 
This situation actually occurs when winter CO$_2$ frost formation rate exceeds
seasonal summer sublimation.
\end{enumerate}

We now put these results in context of an active carbonate-silicate cycle. 
A planet that enters in a global glaciation state must have initially a low CO$_2$ atmospheric content that places it 
in the lower part of the diagram (in Figure~\ref{condens_plots}b). From this point, 
CO$_2$ is outgassed by volcanoes and accumulates in the atmosphere. As the CO$_2$ partial pressure increases, 
the planet moves up in the diagram until: 
\begin{enumerate}
 \item the planet reaches the red zone first. The planet is able to escape glaciation.
 \item the planet reaches the blue zone first. All the extra CO$_2$ possibly outgassed by volcanoes 
condenses at the poles and the planet is locked in a perpetual glaciation state.
\end{enumerate}
For an (Earth-like) 23.5$^{\circ}$ obliquity, 
we find this limit to occur for planets located at more than 1.27~AU from a Sun-like star 
(Flux~$<$~847~W~m$^{-2}$, S$_{\text{eff}}$\footnote{The effective flux S$_{\text{eff}}$ is defined as the ratio between the 
incoming stellar flux on the planet and that on the Earth.}~$<$~0.62), 
with an optimal CO$_2$ partial pressure around 1~bar (see Figure~\ref{condens_plots}b).

However, if a planet starts with a large enough CO$_2$ atmospheric content so that it lies above the blue zone, 
then the CO$_2$ should resist atmospheric collapse and the deglaciation becomes possible.

\subsection{Null obliquity case}
\label{null_obliquity_section}

For a 0$^{\circ}$ obliquity planet, because insolation in the polar regions is lowered, the equator-to-pole temperature contrast is amplified.
Consequently, we find that 
permanent condensation of CO$_2$ at the poles happens for much lower equivalent distances (d~$>$~1.13~AU, 
Flux~$<$~1070~W~m$^{-2}$, S$_{\text{eff}}$~$<$~0.78 ; see Figure~\ref{condens_plots}a). 

\subsection{Effect of CO$_2$ ice clouds}
\label{co2_cloud_section}

During polar nights, the temperatures at the poles are very low. CO$_2$ condenses at the surface but also in the atmosphere, 
forming CO$_2$ ice clouds that may have a significant warming effect \citep{Forg:97}.

We performed multiple simulations for planets at 23.5$^{\circ}$ obliquity including radiatively active CO$_2$ ice clouds.
For this, we used a constant cloud condensation nuclei [CCN] of 10$^5$~kg$^{-1}$ designed to maximize the greenhouse effect of 
CO$_2$ ice clouds \citep{Forg:13}. This idealized configuration corresponds to an endmember description 
of the radiative warming effect of CO$_2$ ice clouds. 
In particular, \cite{Kitz:16} has shown recently, using a more refined radiative transfer scheme, that 
a configuration like the one used in this section probably overestimates the warming effect of CO$_2$ ice clouds.

\begin{figure}
\begin{center}
\label{active_clouds}
\includegraphics[width=9.5cm]{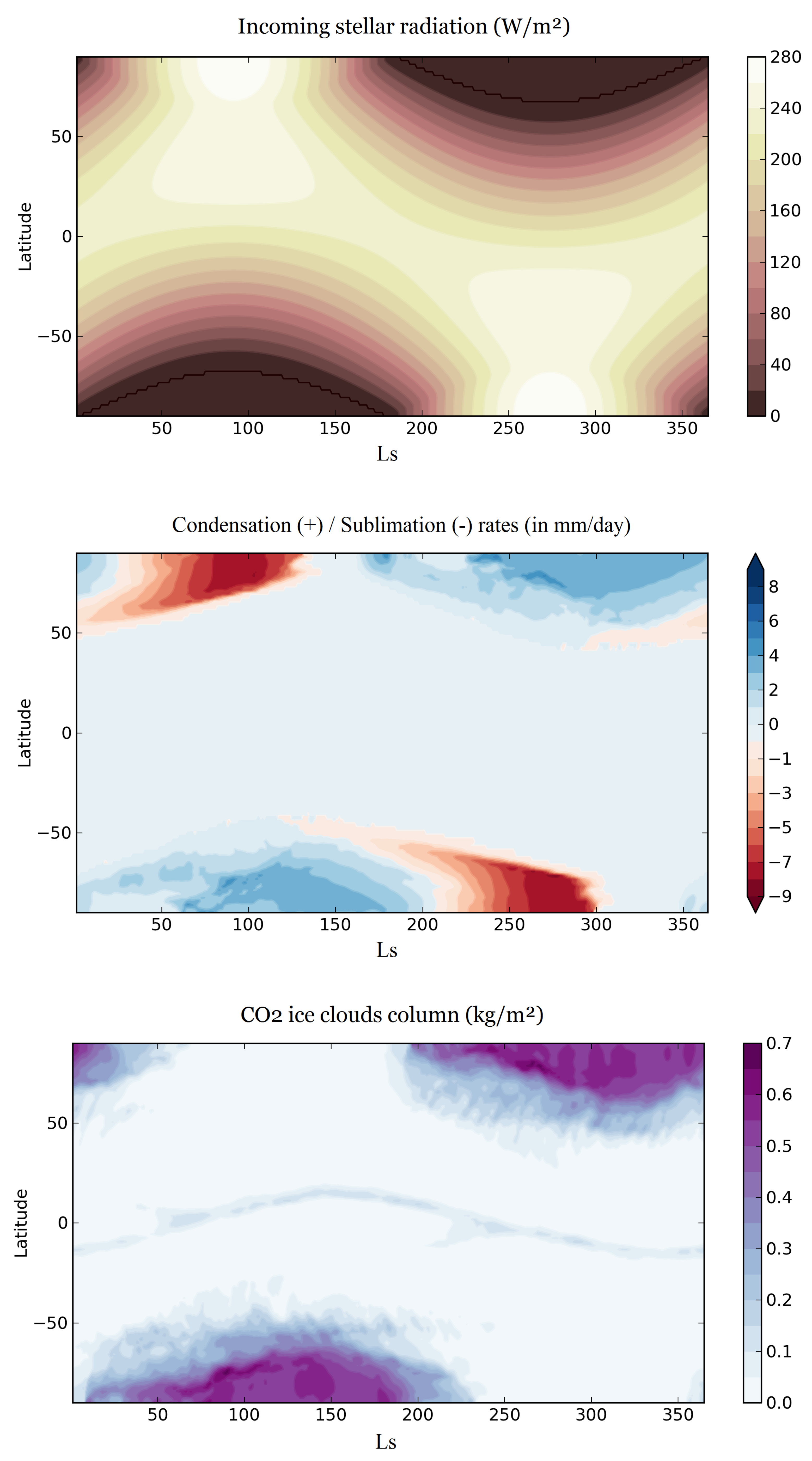}
\caption{From top to bottom: (1) zonal mean of incoming stellar radiation (ISR) versus L$_s$, 
(2) zonal mean of CO$_2$ ice condensation(+)/sublimation(-) rates versus L$_s$, 
and (3) zonal mean of CO$_2$ ice cloud (column integrated) content versus L$_s$.
These figures were computed for a planet with a 23.5$^{\circ}$ obliquity, active CO$_2$ ice clouds, 
located at an equivalent orbital distance of 1.40~AU (F~=~697~W~m$^{-2}$) 
and for a CO$_2$ partial pressure of 1~bar and a N$_2$ partial pressure of 1~bar.}
\end{center}
\end{figure}

Our simulations (example in Figure~\ref{active_clouds}) show that the CO$_2$ ice clouds distribution evolves significantly with seasons.
In winter, CO$_2$ ice clouds form a quasi-complete polar cover and have 
a powerful greenhouse effect that limits the condensation of CO$_2$ at the poles.
In summer, CO$_2$ clouds dissipate as insolation increases. 
As a result, CO$_2$ clouds have a very limited impact on the albedo and 
therefore do not contribute much to prevent polar CO$_2$ ice summer sublimation.

For these two reasons, CO$_2$ ice clouds seriously limit the collapse of CO$_2$ at the poles. Figure~\ref{condens_plots}c shows 
that, with radiatively active CO$_2$ clouds, CO$_2$ condensation would occur for equivalent distance~$>$~1.40~AU 
(Flux~$<$~697~W~m$^{-2}$, S$_{\text{eff}}$~$<$~0.51).

\subsection{Sensitivity study}
\label{sensitivity}

\subsubsection{The albedo of ices}

The choice of water ice albedo can severely affect our results. 
For albedo higher than 0.67, 
we find that CO$_2$ condensation can occur on Earth-like planets (23.5$^\circ$ obliquity, radiatively inactive CO$_2$ ice clouds) for equivalent 
distance as low as 1.15~AU (Flux~$\sim$~1033~W~m$^{-2}$, S$_{\text{eff}}$~$\sim$~0.76), corresponding to the reduced luminosity of the Sun 3.8~Gya 
calculated from standard solar evolution models \citep{Goug:81}.

We note that water ice albedo is considerably reduced around cool stars \citep{Josh:12}, making the scenario of 
CO$_2$ polar condensation less efficient. However, planets orbiting in the Habitable Zone of M-dwarfs 
are also subject to tidal locking. The temperature on the nightside of a synchronous planet can be extremely low, favoring the CO$_2$ 
condensation. This possibility is explored and discussed in details in \citealt{Turbet:2017aa}.

We also explored the effect of CO$_2$ ice albedo that could potentially be very high 
(see section~\ref{ice_properties}, \citealt{Kief:00jgr}) and we found it 
to be much less important.

Eventually, we acknowledge that a real Snowball might have regions of open continent not covered
by water ice, which would lower albedo. These could act as sources of dust, and volcanic aerosols could also darken the surface \citep{Abbot:2010}.

\subsubsection{The rotation rate}

The rotation rate of Earth has evolved in time from 1.2$\times$10$^{-4}$ (4~Gya) 
to 7.3$\times$10$^{-5}$~rad~s$^{-1}$ (now) due to the effect of the Moon's tidal friction \citep{Walk:86,Zahn:87}. 
More generally, Earth-like planets could harbor a wide range of rotation rates.

In this study, we find that rotation rate is a very important parameter for CO$_2$ polar condensation. 
A high rotation rate reduces the latitudinal transport and thus increases the equator-to-pole 
surface temperature gradient \citep{Kasp:15}, favoring CO$_2$ condensation at the poles.
We find that, for a rotation rate $\Omega$~=~4~$\Omega_{\text{Earth}}$ (23.5$^\circ$ obliquity, radiatively inactive CO$_2$ ice clouds), 
CO$_2$ can collapse as early as 1.15~AU (Flux~$\sim$~1033~W~m$^{-2}$, S$_{\text{eff}}$~$\sim$~0.76).

To go further, we investigated the case of archean Earth, 3.8~Gya (S$_{\text{eff}}$~$\sim$~0.76), 
through simulations of a completely frozen planet, assuming 
a 23.5$^{\circ}$ obliquity, a rotation period of 14~hours and a water ice albedo set to 0.6. We find that CO$_2$ can condense at the poles seasonally, 
but never permanently.

\subsubsection{The planetary mass}

Massive planets also have a large radius that may 
be responsible for the weakening of the poleward heat transport, favoring the collapse of CO$_2$. 
For example, assuming an internal composition similar to the Earth, a 5~M$_{\text{Earth}}$ planet would have 
a radius around 1.5~R$_{\text{Earth}}$ or 10$^4$~km \citep{Seager:2007}.
We performed GCM simulations for a 5~M$_{\text{Earth}}$ planet, for a CO$_2$ partial pressure of 1~bar and a N$_2$ partial pressure of 1~bar. 
In these conditions, we find that the collapse of CO$_2$ can occur 
for equivalent distances as low as 1.18~AU (Flux~$<$~981~W~m$^{-2}$, S$_{\text{eff}}$~$<$~0.72).

However, because accretion of volatiles should increase with planetary mass,
the atmospheric mass should also increase with it (see \cite{Kopp:14} section~2 for more details). 
Therefore, we expect a massive planet to have a greater content 
of background gas (e.g. N$_2$). For a 5~M$_{\text{Earth}}$ planet, assuming that P$_{\text{surface}}$ scales as {R$_{\text{planet}}$}$^{2.4}$.
(equation~3 in \cite{Kopp:14}), we roughly get a partial pressure of N$_2$ equal to 3.3~bars
\footnote{We remind the reader that volatile delivery to a planet is stochastic in nature and may
be a weak function of planetary mass. By way of comparison, Venus (at $\sim$~0.8~M$_{\text{Earth}}$) has $\sim$~3 bars of N$_2$ in its atmosphere.}.
Using GCM simulations, we find that in this more realistic case, CO$_2$ collapses for equivalent distances 
greater than 1.30~AU (Flux~$<$~808~W~m$^{-2}$, S$_{\text{eff}}$~$<$~0.59).

\subsubsection{The atmospheric composition: the role of N$_2$ partial pressure}

The result of the previous section shows that N$_2$ partial pressure can have a significant impact on CO$_2$ collapse.
As pN$_2$ increases, the atmosphere thickens and (1) the poleward heat transport increases and (2) the greenhouse effect of CO$_2$ increases 
because of pressure broadening of CO$_2$ by N$_2$. Both effects tend to prevent CO$_2$ condensation.

Quantitatively, our simulations show that, for an Earth-like planet with a CO$_2$ partial pressure of 1~bar (with 23.5$^\circ$ obliquity, 
radiatively inactive CO$_2$ ice clouds), 
doubling pN$_2$ from 1 to 2~bars leads to the condensation of CO$_2$ for equivalent orbital distances 
greater or equal to 1.35~AU (+0.08~AU difference).

\subsubsection{The topography}
\label{topography}

To investigate the possible effects of topography on CO$_2$ polar condensation, we ran several simulations in which 
we emplaced a mountain range similar in size to the Himalaya (5000~km altitude plateau, 10$^6$~km$^2$) at several latitudes 
(0$^\circ$N, 30$^\circ$N, 45$^\circ$N, 60$^\circ$N, 90$^\circ$N). We find that CO$_2$ permanent condensation  
can occur on top of the mountain range for distances lower than 1.27~AU, 
only for latitudes greater than 45$^\circ$N. For the 60$^\circ$N latitude case, CO$_2$ condensation 
starts as low as 1.20~AU (Flux~$\sim$~949~W~m$^{-2}$, S$_{\text{eff}}$~$\sim$~0.69).

Nonetheless, the total amount of CO$_2$ that could condense and be trapped above a mountain range is rather low, 
because it is limited by the area of the mountain range.

\section{How much CO$_2$ ice can be trapped ?}

In this section, we investigate two processes that should control the amount of CO$_2$ that can be trapped on the surface 
or subsurface: 1) flows of polar CO$_2$ ice to lower latitudes and
sublimation; and 2) burial of CO$_2$ ice beneath the water ice cover due to higher density.

\subsection{Maximum size of CO$_2$ ice glaciers}

At first sight, the main limit of the trapping of CO$_2$ as ice instead of greenhouse gas 
is the size of the solid CO$_2$ polar reservoirs. When CO$_2$ is outgassed by volcanoes, 
the atmospheric pressure stays constant but the size of the CO$_2$ polar caps grows. At some point, the CO$_2$ ice caps form glaciers 
that can flow efficiently toward equatorial regions, CO$_2$ ice being much less viscous than water ice. 
In the process, CO$_2$ ice can be sublimated and reinjected in the atmosphere. 

In this section, we give estimates of the 
maximum amount of CO$_2$ that can be trapped in steady state CO$_2$ ice polar caps, for a planet with a 23.5$^{\circ}$ obliquity, 
radiatively inactive CO$_2$ ice clouds, located at a distance of 1.30~AU (F~=~808~W~m$^{-2}$) 
and for both CO$_2$ and N$_2$ partial pressures set to 1~bar.

\subsubsection{CO$_2$ ice caps radial extent}
\label{co2_rad_extent}

First, we get from our GCM simulations the radial extent of the two permanent CO$_2$ ice polar caps, defined for 
a positive annual mean CO$_2$ condensation rate. We note the corresponding radius of the ice cap $R_{\text{cap,} 1}$.
Then, in GCM simulations, we artificially extend (in the radial direction) 
the CO$_2$ ice caps to take into account the glacier flow, 
until the net globally averaged annual condensation$/$sublimation rate vanishes.
In practice, we used the following algorithm:
\begin{enumerate}
 \item We run the simulation until globally averaged annual mean rate of condensation/sublimation is constant.
 \item If this rate is (roughly) null, CO$_2$ ice caps are in a dynamical equilibrium. We stop here.
 \item If this rate is positive, we artificially increase the size of the CO$_2$ ice caps of one GCM grid (in the latitudinal direction, and 
for each longitude) by emplacing arbitrarily a large enough amount of CO$_2$ ice. We go back to step~1.
\end{enumerate}
This method gives us a good estimate of the steady state maximal 
extent of the CO$_2$ ice polar caps. We note the corresponding radius of the ice cap $R_{\text{cap,} 2}$.
\begin{figure}
\begin{center}
 \includegraphics[width=10cm]{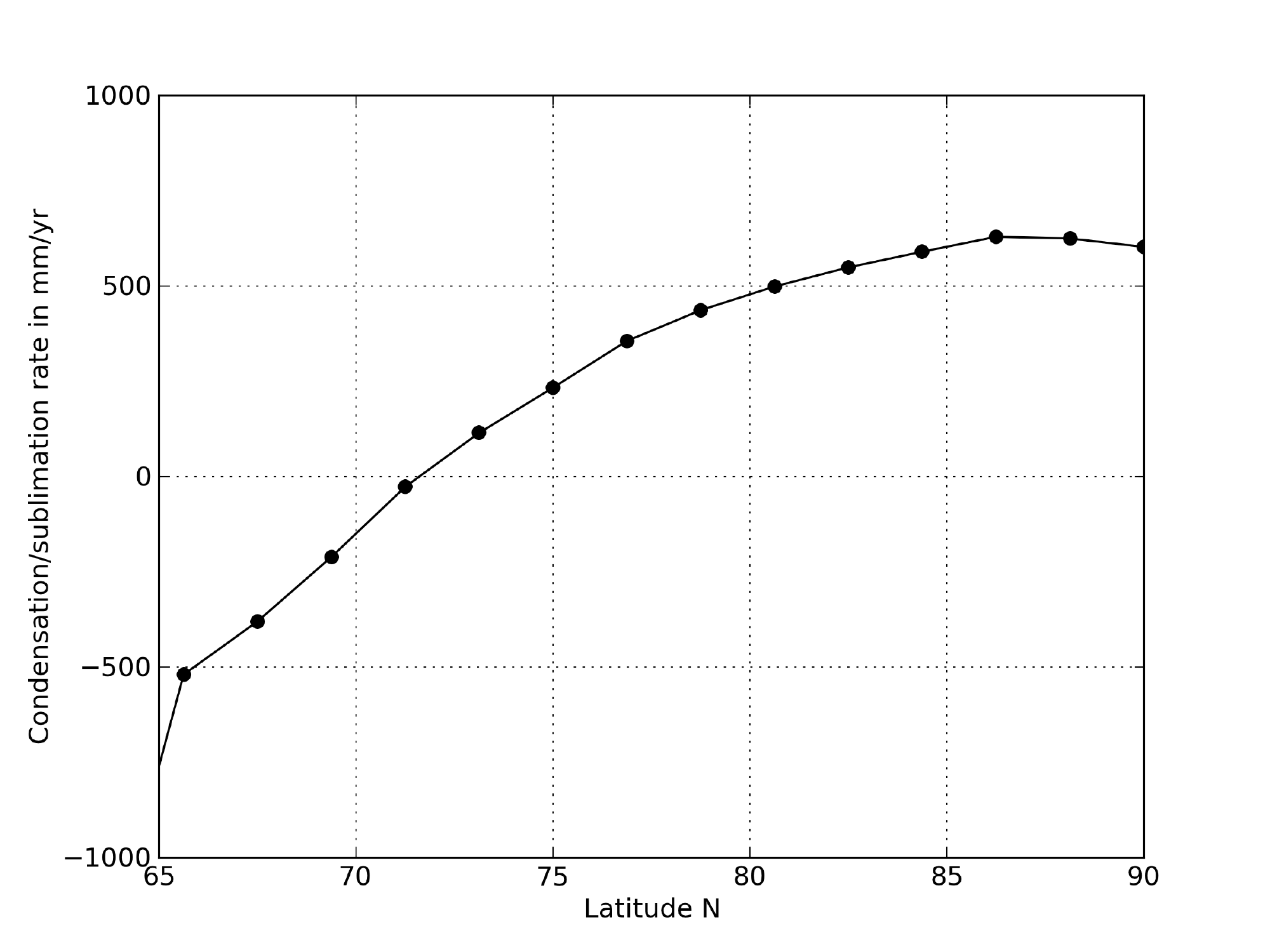}
\caption{Zonal and annual mean CO$_2$ condensation/sublimation rates along the CO$_2$ northern ice cap, for the reference 
case (1.30~AU equivalent distance, 23.5$^{\circ}$~obliquity and radiatively inactive CO$_2$ ice clouds). 
It corresponds also to the a(r) function (see section~\ref{noslip}). 
Positive values correspond to latitudes where the annual mean CO$_2$ condensation rate is positive. 
These are locations where CO$_2$ condenses permanently.
Negative values are "potential" sublimation rates, i.e. it is assumed that CO$_2$ ice is available -- 
following the algorithm described in section~\ref{co2_rad_extent} -- and can sublime all year long. 
}
\label{zonal_condensation}
\end{center}
\end{figure}
With and without glacier flow, we obtain polar cap radii $R_{\text{cap,} 1}$~=~2070 km and $R_{\text{cap,} 2}$~=~2800 km. 
This corresponds to a latitudinal extent down to 71.5$^{\circ}$N and 65$^{\circ}$N
\footnote{By symmetry, results are identical in the southern hemisphere.}, respectively. 
Figure~\ref{zonal_condensation} shows zonal condensation/sublimation annual mean rates along the North Pole 
CO$_2$ glacier.

In total, the two CO$_2$ ice caps have an area of 27~millions of km$^2$ (respectively 49~millions of km$^2$ if considering the glacier flow). 
This corresponds to 5$\%$ (respectively 10$\%$) of the total area of the (Earth-like) planet.

\subsubsection{CO$_2$ ice cap inner region: a thickness limited by basal melting}

In the region within the CO$_2$ ice caps with positive annual mean condensation rates, 
the CO$_2$ ice maximum thickness should be mainly limited by basal melting induced by the geothermal heat flux, noted $F_{\text{geo}}$.

\begin{figure}
\begin{center}
 \includegraphics[width=13cm]{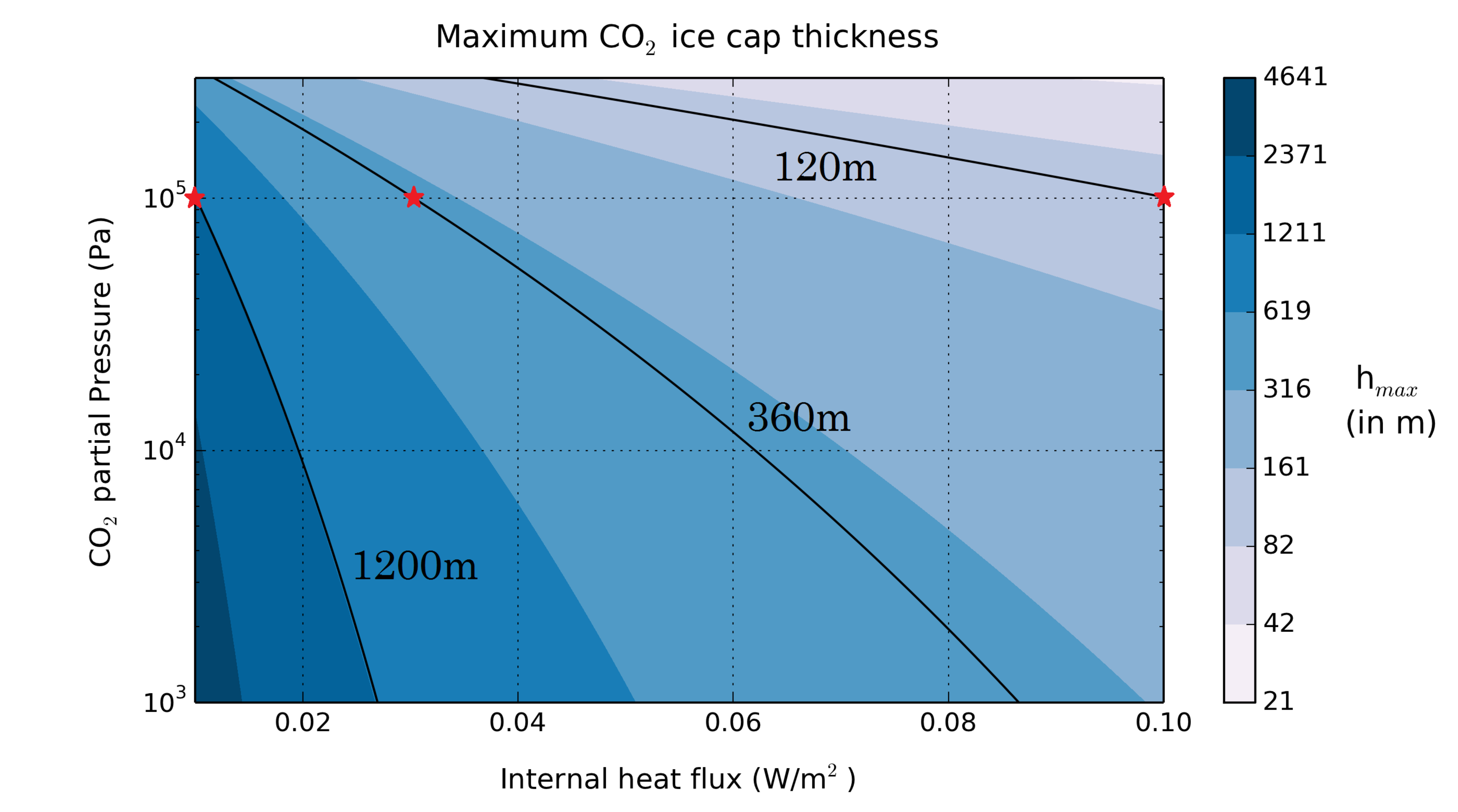}
\caption{CO$_2$ ice cap maximum thickness (in m) calculated from the basal melting condition,
as a function of surface CO$_2$ partial pressure and internal heat flux. 
The red stars correspond to three distinct cases, with pCO$_2$~=~1bar and $F_{\text{geo}}$~=~10~/~30~/~100~mW~m$^{-2}$, 
that are investigated in more details in the following sections.}
\label{2D_basal_melting}
\end{center}
\end{figure}

Assuming that the temperature inside the CO$_2$ glacier rises linearly with depth, 
with a lapse rate fixed by the geothermal heat flux (conductive regime), the maximum CO$_2$ ice thickness $h_{\text{max}}$ is given by: 
\begin{equation}
 h_{\text{max}}~=~\frac{\lambda_{\text{CO}_2}~(T_{\text{melt}}-T_{\text{surf}})}{F_{\text{geo}}},
\label{conductivity}
\end{equation}
where $\lambda_{\text{CO}_2}$ is the thermal conductivity of CO$_2$ ice, $T_{\text{surf}}$ is the temperature at the top of the glacier and 
$T_{\text{melt}}$ is the melting temperature of CO$_2$ ice at the base of the glacier.
As the latter is a function of the pressure below the CO$_2$ glacier, it implicitly depends on $h_{\text{max}}$ 
so that the above equation must be solved numerically, once the local surface temperature is known 
(see Appendix~\ref{max_ice_basal_melting} for details on the calculations).
The resulting maximum CO$_2$ ice cap thickness $h_{\text{max}}$ is plotted on 
Figure~\ref{2D_basal_melting} as a function of the internal heat flux and the CO$_2$ partial pressure.
For geothermal heat fluxes of 100/30/10~mW~m$^{-2}$ (red stars in Figure~\ref{2D_basal_melting}), 
basal melting condition gives CO$_2$ ice maximum thicknesses of 120/360/1200~m. 

The exact latitude at which the glacier maximum thickness is determined by either basal melting condition 
or by glacier flow is difficult to determine. Moreover, in the region of transition between these two regimes, a basal liquid CO$_2$ flow could 
persist\footnote{Such basal flow could carry heat away from basal melting regions and thus also reduce the heat 
flux conducted through the CO$_2$ ice caps \citep{Cuff:10,Meno:13}.} and therefore speed up the glacier flow.

For simplicity, we fixed the thickness of the steady-state CO$_2$ glacier in the regions 
with positive annual mean condensation rate (regions of latitude~$>$~71.5$^{\circ}$N ; see Figure~\ref{zonal_condensation})
at the constant maximum thickness h$_{\text{max}}$ derived from the basal melting condition. 
This maximum thickness serves as a boundary condition for the calculation of the glacier flow in regions with 
positive annual mean sublimation rate (see next section).

\subsubsection{CO$_2$ ice cap outer region: a thickness limited by glacier flow and sublimation}
\label{noslip}

To get estimates of CO$_2$ glacier profiles in the regions where sublimation dominates over condensation 
(latitude~$<$~71.5$^{\circ}$N, see Figure~\ref{zonal_condensation}), we use a simple isothermal glacier model
\footnote{This is a simple approach that does not physically describe how 
the accumulating CO$_2$ ice spreads to feed the ablation zone. We acknowledge that 
a more physically complete model should be developed in the future.} 
(assuming no slip at the base of the glacier), 
following a similar approach than \cite{Meno:13}.
The steady-state equation 
satisfied by a thin isothermal CO$_2$ glacier with a flat base and a no
slip boundary condition at its bottom is \citep{Fowl:11,Paterson:2010}\footnote{This is the classic solution for a glacier shape.}:
\begin{equation}
 \frac{1}{r}\frac{\partial}{\partial r} \Bigg(r~ \frac{2 A_0 {(\rho g)}^n}{n+2}~h^{n+2}~{\mid \frac{\partial h}{\partial r}\mid}^{n-1}~\frac{\partial h}{\partial r} \Bigg)~+~a(r)~=~0, 
\label{glen_law}
\end{equation}
with $h(r)$ the thickness of the CO$_2$ ice cap, assumed (for simplicity) radially symmetric, 
A$_0$ the flow rate constant (in Pa~s$^{-n}$), 
n the power-law creep exponent, 
g the surface gravity, 
$\rho$ the density of CO$_2$ ice and 
a(r) the annual mean condensation rate assumed (also for simplicity) to be a function of r only, 
and derived from our GCM simulations (see Figure~\ref{zonal_condensation}).

First, assuming $\frac{\partial h}{\partial r}~=~0$ for $r~=~R_{\text{cap,1}}$, we integrate equation~\ref{glen_law} once.
Then, assuming $h~>~0$, $\frac{\partial h}{\partial r}~<~0$ and $a(r)<0$, we separate the variables $h$ and $r$, 
and we integrate a second time, assuming $h$($R_{\text{cap,2}}$)~=~$0$ at the edge of the glacier.
Eventually, assuming that $h$($R_{\text{cap,1}}$)~=~$h_{\text{max}}$, with $h_{\text{max}}$ the maximum thickness 
of the glacier calculated using the basal melting condition, we can constrain the flow rate A$_0$ and get 
the following solutions for the thickness $h$ of the CO$_2$ ice glacier:

\begin{equation}
h(r)~=~h_{\text{max}}\text{  for }r \leqslant R_{\text{cap,1}},
\label{solution_glacier1}
\end{equation}

\begin{equation}
h(r)~=~h_{\text{max}}\Bigg( \frac{\int_{r}^{R_{\text{cap,2}}} {\Big( -\frac{1}{r_2}~\int_{R_{\text{cap,1}}}^{r_2} r_1~a(r_1)~dr_1 \Big)}^{\frac{1}{n}}dr_2}
{\int_{R_{\text{cap,1}}}^{R_{\text{cap,2}}} {\Big( -\frac{1}{r_2}~\int_{R_{\text{cap,1}}}^{r_2} r_1~a(r_1)~dr_1 \Big)}^{\frac{1}{n}}dr_2} \Bigg)^{\frac{n}{2n+2}}
\text{  for }r>R_{\text{cap,1}}, 
\label{solution_glacier2}
\end{equation}

\begin{figure}
\begin{center}
 \includegraphics[width=12cm]{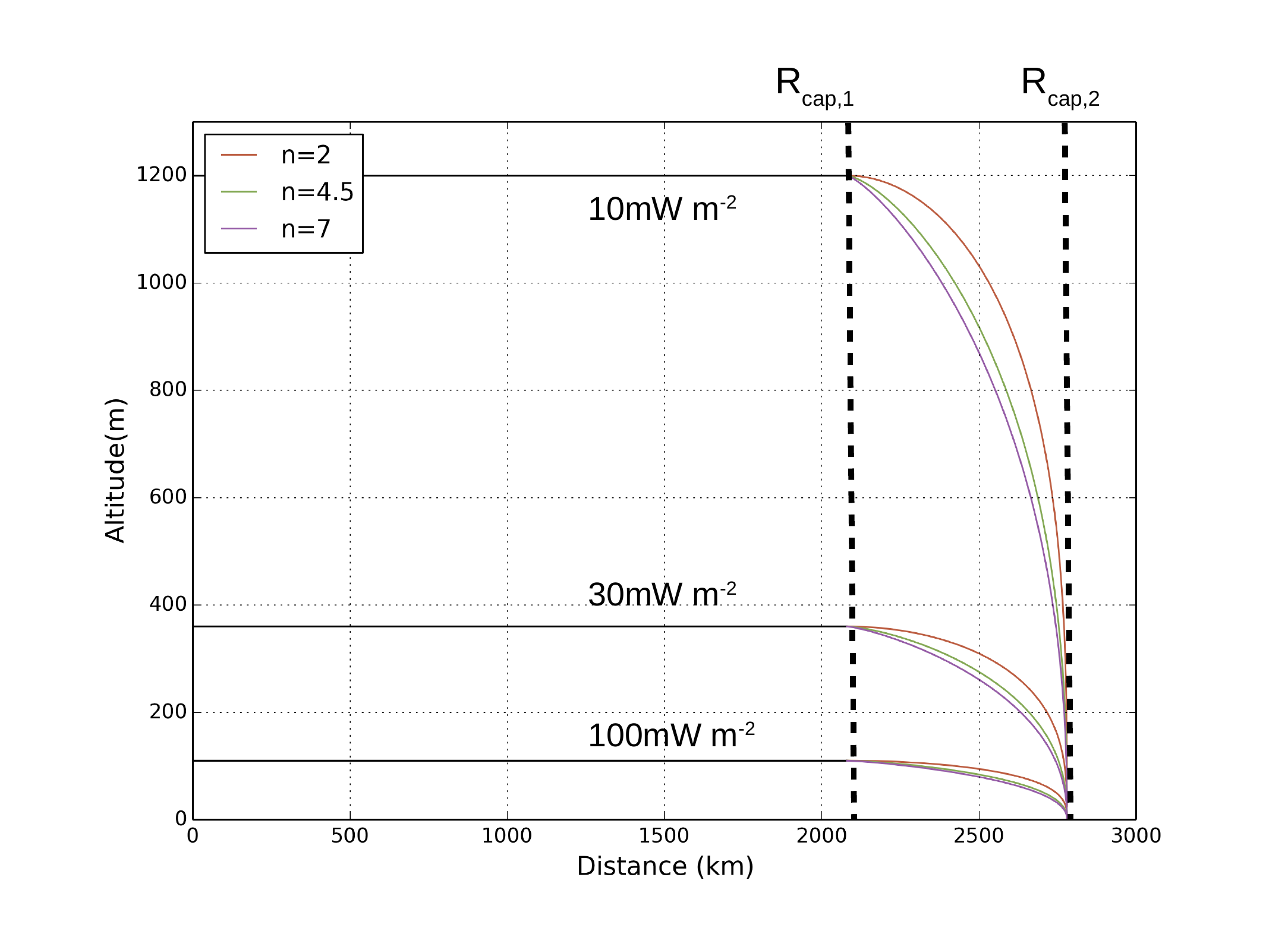}
\caption{Possible radial profiles for the CO$_2$ ice polar glaciers. 
They correspond to solutions of equations \ref{solution_glacier1} and \ref{solution_glacier2}, for 
3 different internal heat fluxes (10/30/100~mW~m$^{-2}$) and three different flow laws (n~=~1, 4.5 and 7).}
\label{glacier_profile}
\end{center}
\end{figure}

Figure~\ref{glacier_profile} presents 9 possible steady-state CO$_2$ ice glacier profiles, 
for 3 different geothermal fluxes and 3 different flow laws (n~=~1, 4.5 and 7) chosen consistent with previous 
experimental studies \citep{Durh:99} and works about the martian CO$_2$ southern polar cap \citep{Nye:00}.

\subsubsection{Calculation of the CO$_2$ maximum total reservoir}
\label{max_reservoir_co2}

After integration of the steady state CO$_2$ ice radial profiles, 
we find that the maximum amount of CO$_2$ ice that can be stored in the two polar caps
is approximately equal to 1.5/4.5/15~bars\footnote{These quantities are not affected much by the choice of flow law (see Figure~\ref{glacier_profile}).} 
for geothermal heat fluxes of 100/30/10~mW~m$^{-2}$, respectively.
On Earth, the current geothermal heat flux is $\sim$~80~mW~m$^{-2}$ but can varies a lot, 
typically from 20~mW~m$^{-2}$ up to 400~mW~m$^{-2}$, depending on the regions \citep{Davi:13}.
Approximately 60$\%$ (respectively 40$\%$) of CO$_2$ ice is 
trapped in region of $r<R_{\text{cap,1}}$ (respectively $r>R_{\text{cap,1}}$), meaning that 
the accumulation zone and the ablation zone contribute significantly 
and equally to the total size of the CO$_2$ ice reservoir.

It is now possible to calculate, for a given rate of CO$_2$ volcanic outgassing, the time required for a hard snowball planet to 
fill the CO$_2$ ice polar reservoirs and thus potentially escape glaciation.
Assuming a present-day Earth CO$_2$ volcanic outgassing rate of $\sim$~7~teramoles/year or 60~bars/Gy 
\citep{Bran:95,Jarr:03}, we get durations of glaciation periods of $\sim$~42/93/270~My for 
internal heat fluxes of 100/30/10~mW~m$^{-2}$. In fact, because CO$_2$ volcanic outgassing rate is dependant on the 
internal heat flux, the duration of the glaciation phases could be drastically increased for planets with low geothermal 
heat fluxes (typically lower than 80~mW~m$^{-2}$ as on present-day Earth). 
This suggests that planets near the outer edge of the habitable zone 
may need to be quite volcanically active in order to remain habitable.

When (and if) CO$_2$ ice caps become full, the system suddenly reaches a runaway phase, 
ending up with the complete vaporization of CO$_2$ polar ice caps and thus a hot atmosphere.
This may have implications in the calculation and description of 'limit-cycles' \citep{Meno:15,Haqq:16,Abbo:16} and more 
generally on the habitability of distant Earth-like planets. Detailed calculations of 
the stability of CO$_2$ ice caps when full can be found in Appendix~\ref{stability_co2}.

It is also conceivable that catastrophic events such as meteoritic impacts could trigger deglaciation.

\subsection{Gravitational stability and CO$_2$ sequestration}

For planets covered by several hundred meters of water ice\footnote{We remind here that the Earth Global Equivalent Layer of surface water 
is $\sim$~3km.}, owing to the high density of CO$_2$ ice ($\sim$1500~kg~m$^{-3}$) 
or liquids ($\sim$1170~kg~m$^{-3}$) compared to water ice ($\sim$930~kg~m$^{-3}$ at 200-220~K), 
the CO$_2$ deposits should rapidly become gravitationally unstable. 
This would result in a burial of the CO$_2$ deposits at the base of the H$_2$O ice layer. 
Such a density contrast should initiate Rayleigh-Taylor instabilities at a timescale that can be estimated 
at first order assuming two isoviscous layers using the following relationship \citep{Turc:01}:

\begin{equation}
\tau_{\text{RT}}=\frac{13~\eta}{\Delta\rho g b},
\label{eq_turcotte}
\end{equation}
with $\eta$ the viscosity of the more viscous layer, $\Delta\rho$ the density contrast between the two layers 
and $b$ the characteristic size of the domain. 
Depending whether CO$_2$ is liquid or solid at the interface with the H$_2$O layer, the density contrast would range between 240 and 570~kg~m$^{-3}$. 
For a CO$_2$ ice deposit thickness of about 100~m, this density contrast results in a stress at the 
interface between the two layers of the order of 1~MPa. For such a stress level and temperature 
ranging between 200 and 220~K, the viscosity of the CO$_2$ ice layer is estimated between 10$^{12}$-10$^{13}$ Pa~s, 
based on available experimental data \citep{Durh:99}.
At the same temperature and stress conditions, water ice has a viscosity $\le$ 10$^{16}$ Pa~s, for grain size lower than 1 mm, based
on experimental data \citep{Durh:01,Durh:01AREPS,Gold:01}. Therefore, this would be the layer controlling the Rayleigh-Taylor timescale. 
Assuming that a thickness of 100~m of CO$_2$ deposit is representative of the characteristics size of the domain, 
the R-T timescale is 7$\times$10$^3$~yrs~$\times$~($\eta$/10$^{16}$ Pa s), which is geologically short. 
In the case of basal melting of the CO$_2$ layer, 
we should keep in mind that the generated CO$_2$ liquids will be unstable relative to the underneath water ice layer, 
but also relative to the overlying CO$_2$ ice layer as it has a density intermediate between those of CO$_2$ ice and water ice. 
When CO$_2$ melting occurs at the base of the CO$_2$ layers, 
a competition is expected between downward drainage through 
the water ice layer (similar to crevasse hydrofracturing in Earth’s 
glaciers; \citealt{Kraw:09}) and liquid injection in the overlying CO$_2$ ice 
layer and refreezing (similar to dike propagation). Assessing in details such complex 
processes are out of scope of the present paper. Nevertheless, based on these considerations, 
we can safely argue that the accumulation of a large volume of liquid CO$_2$ at the base of the CO$_2$ ice layer is unlikely. 

Once gravitationally destabilized, 
the CO$_2$ ice deposit would sink at the base of the water ice shell 
at a rate that is determined mostly by the viscosity of water ice 
and the size of the CO$_2$ ice diapir. The time required for a CO$_2$ ice 
diapir to cross the water ice layer can be estimated using the Stokes 
velocity, the terminal velocity of a sphere falling through 
a constant viscosity medium \citep{Ziet:07}: 
\begin{equation}
U_s~=~\frac{2}{9}~\Delta\rho g~(r^2 / \eta)
\label{eq_stokes}
\end{equation}
For a diapir radius $r$ of 100~m (comparable to thickness of the CO$_2$ deposit) 
and a viscosity of water ice of 10$^{15}$-10$^{16}$ Pa~s, this leads to a velocity of 0.04-0.4~m$/$yr. 
As temperature increases as a function of depth, the viscosity of water ice 
is expected to decrease with depth, resulting in an acceleration of the diapir fall. 
A 100-m diapir of CO$_2$ ice would thus not need more than $\sim$~10$^4~$yrs to reach the bottom 
of a 1-km thick water ice layer. 

Even if the CO$_2$-H$_2$O interface is beneath the melting temperature 
of CO$_2$ ice, melting may occur during the diapir fall, which 
will reduce the density contrast and hence the fall velocity. However, 
as mentioned above, melting may promote fracturing of the water ice medium and 
rapid drainage of the CO$_2$ liquids due to its high density relative to the 
surrounding water ice. The rate of melting should also depend on the efficiency of heat 
transfer between the diapir and the surrounding ice. As the thermal diffusive timescale ($\sim$r${^2/\kappa}$) 
for a 100-m-size diapir of CO$_2$ is of the order of 10$^3$~yrs and is thus $\le$ to the expected diapir fall timescale, 
melting during the descent would be efficient. Detailed modelling would be required though to determine the exact volume of generated melt during descent.

CO$_2$ ice should completely melt and equilibrate thermally with the surrounding H$_2$O 
media once stabilized at the bottom of the water ice shell. 
The temperature and pressure conditions at the bottom of the water ice layer depend on its thickness and on the 
geothermal flow. For geothermal heat flux between 50 and 100~mW~$^{-2}$, typical of the Earth, the melting of water ice would be 
reached for depth ranging between 1.5 and 3~km, corresponding to pressure of about 14 and 28~MPa. 
This pressure range is well above the saturation vapor pressure (3.5~MPa at 273~K; \citealt{Lide:04}), 
so that CO$_2$ is highly stable in the liquid phase. Destabilizing the CO$_2$ liquids 
into gas phase would require the 273~K-isotherm 
at a depth of only 375~m, corresponding to abnormally high geothermal heat flux of 400~mW~$^{-2}$, 5 times larger than 
the average heat flux on Earth. Even if the density of liquid CO$_2$ decreases with increasing temperature as it equilibrates 
with the surrounding water ice media, it remains always denser than water ice \citep{Span:96}, and therefore should always accumulate 
at the bottom of the ice shell. At T~=~273~K and pressure between 14 and 28~MPa, CO$_2$ liquids have a density very close to that of 
liquid water (994 and 1048 kg~m$^{-3}$, respectively, using the equation of state of \cite{Span:96}), so that CO$_2$ coexists with 
H$_2$O at the ice-water interface.

Pressure and temperature conditions expected at the bottom of the ice layer are in the stability field of CO$_2$ clathrate hydrate 
\citep{Sloa:98,Long:05}, therefore CO$_2$ should rapidly interact with H$_2$O molecules to form clathrate hydrate. 
Clathrate hydrates are non-stoichiometric compounds consisting of hydrogen-bonded H$_2$O molecules forming cage-like structures 
in which guest gas molecules, such as CO$_2$, can be trapped \citep{Sloa:98}. Once formed, these clathrates are very stable and can 
be dissociated only if the temperature is raised about 5-10~K above the melting point of water ice. The storage of CO$_2$ in the form of 
clathrate should be particularly efficient in the case of warm bottom conditions as CO$_2$ liquids and liquid water coexist. In the cold 
bottom condition with no water ice melting, clathration is still possible but would be less efficient due to the physical separation between the 
two components. As CO$_2$ clathrate hydrate has a density of about 1150 kg~m$^{-3}$ (assuming full cage occupancy, \cite{Sloa:98}), they would rapidly 
sink at the bottom of the water liquid layer, ensuring an almost complete clathration of CO$_2$. Part of the CO$_2$ should dissolve in the liquid 
water during the clathrate sinking. The relative proportion of CO$_2$ trapped in the form of clathrate hydrate or dissolved in the water layer 
would depend on the volume of CO$_2$ that is buried at the base of the ice shell and on the volume (thickness) of the water layer. 

In summary, as long as the water ice shell exceeds a few hundred meters, CO$_2$ should remain sequestered either in the form of CO$_2$ liquids, 
in the form of CO$_2$ clathrate hydrate or dissolved in the subglacial water layer. Release of gaseous CO$_2$ may occur only due to an increase 
of surface insulation and increase of geothermal flux resulting in a significant thinning and breaking-up of the water ice shell.   
The total amount of CO$_2$ that can be stored in the H$_2$O layer (by any of these three processes) depends on the total abundance of H$_2$O of 
the planet as well as CO$_2$/H$_2$O ratio. 
The CO$_2$ inventory on Earth, mostly in the form of carbonate rocks on the continents, 
is of the order of 10$^2$~bars \citep{Walker:1985}. If the total CO$_2$ inventory exceeds the total amount of CO$_2$ that can be stored in the 
water layer, then the planet should be able at some point to escape from global glaciation, 
extra CO$_2$ being likely returned to the atmosphere through cryovolcanism.

Evaluating the maximum amount of CO$_2$ that can be trapped underneath the water 
ice cover require however a detailed description of the H$_2$O layer structure as well 
as thermodynamic models predicting the partitionning of CO$_2$ between the different phases.

\section{Conclusions}

We highlight in this paper a new scenario that would prevent distant Earth-like planets 
orbiting Sun-like stars from escaping episodes of complete glaciation.
When a terrestrial planet reaches a Snowball state, CO$_2$ weathering ceases and CO$_2$ can accumulate in the atmosphere because 
of volcanic outgassing. As CO$_2$ builds up in the atmosphere, 
the temperature of condensation of CO$_2$ can exceed the surface temperature of the poles, which leads to the 
trapping of all extra CO$_2$ possibly outgassed. 

Using LMD Global Climate Model simulations designed for Earth-like fully glaciated planets, 
we show that this mechanism can work from orbital distances as low as 1.27~AU 
(Flux~$\sim$~847~W~m$^{-2}$, S$_{\text{eff}}$~$\sim$~0.62). By comparison, the most recent estimate of the 
outer edge of the Habitable Zone (e.g. the maximum greenhouse limit) in similar conditions is 1.67~AU 
(Flux~$\sim$~490~W~m$^{-2}$, S$_{\text{eff}}$~$\sim$~0.36) \citep{Kopp:13}.

This limit can occur at even lower distances for planets with 1) a low obliquity, 
2) a low N$_2$ partial pressure, 3) a high rotation rate and 4) a high value of 
the water ice albedo (we chose in this work the conservative value of 0.6).

Conversely, this limit can occur at significantly higher distances when taking into account the radiative effect of CO$_2$ ice clouds. 
In winter, CO$_2$ ice clouds that form at the cold poles scatter back thermal infrared radiation of the surface. In summer, 
clouds are dissipated and therefore have almost no impact on the bond albedo. For these two reasons, 
CO$_2$ ice clouds can have a powerful warming effect at the poles, limiting CO$_2$ collapse at orbital 
distances greater than 1.40~AU (Flux~$<$~697~W~m$^{-2}$, S$_{\text{eff}}$~$<$~0.51).

For each possible configuration, the amount of CO$_2$ that can be trapped 
in polar CO$_2$ ice caps depends on the efficiency of CO$_2$ ice to flow laterally as 
well as its graviational stability relative to subsurface water ice. 
The flow of CO$_2$ ice from polar to low latitudes regions is mostly 
controlled by its basal temperature (due to both the conductivity and melting temperature of CO$_2$ ice being low), 
and hence by the internal heat flux of the planet.
We find for example that a frozen Earth-like planet located at 1.30~AU of a Sun-like star could store
as much as 1.5/4.5/15~bars of CO$_2$ ice at the poles, for internal heat fluxes of 100/30/10~mW~m$^{-2}$. 

But these amounts are in fact lower limits. 
CO$_2$ ice being denser than water ice ($\sim$~1.7~$\times$ the volumetric mass of water ice), 
we find that CO$_2$ ice deposits should be gravitationally unstable and get buried beneath the water ice cover 
in a geologically short timescale of $\sim$~10$^4$~yrs, mainly controlled by 
the viscosity of water ice. If water ice cover exceeds 
about 300~meters (e.g. 10$\%$ of the Earth hydrosphere), then CO$_2$ should be permanently sequestered underneath the water 
ice cover, in the form of CO$_2$ liquids, CO$_2$ clathrate hydrates and/or dissolved in subglacial 
water reservoirs (if any). This would considerably 
increase the amount of CO$_2$ trapped and further reduce the probability of deglaciation.

\section{acknowledgements}
M.T. thanks Emmanuel Marcq for fruitful discussions related to this work during the legendary LMD PLANETO meetings. 
We acknowledge the feedbacks from Jim Kasting, William B. McKinnon, Dorian Abbot and an anonymous referee, 
which improved both the style and content of the manuscript.
This project has received funding from the European Research Council (ERC) under 
the European Union’s Horizon 2020 research and innovation programme (grant agreement No. 679030/WHIPLASH).

\appendix

\section{The LMD Generic Global Climate Model}
\label{appen_lmd_model}

\subsection{Model core}

This model originally derives from the LMDz 3-dimensions 
Earth Global Climate Model \citep{Hour:06}, 
which solves the primitive equations of geophysical fluid dynamics using 
a finite difference dynamical core on an Arakawa C grid.

The same model has been used to study many 
different planetary atmospheres of low-irradiated planets, 
including Archean Earth \citep{Char:13}, 
past climates of Mars \citep{Forg:13,Word:13,Turbet:2017icarus}, 
or exoplanets \citep{Word:11ajl,Turbet:2016aa, Turbet:2017aa}.

The simulations presented in this paper were all performed at 
a horizonal resolution of 96~$\times$~96 (e.g. 3.75$^{\circ}$~$\times$~1.875$^{\circ}$) in longitude~$\times$~latitude.
In the vertical direction, the model is composed of 15 distinct atmospheric layers that were 
designed using hybrid $\sigma$ coordinates (where $\sigma$ is the ratio between 
pressure and surface pressure).

To account for thermal conduction in the icy ground, we used a 18-layers thermal diffusion soil model. 
Mid-layers depths range from d$_0$~$\sim$~0.1~mm to d$_{17}$~$\sim$~18~m, following the power law 
d$_n$~=~d$_0$~$\times$~2$^n$ with $n$ being the corresponding soil level, chosen to take 
into account both diurnal and seasonal thermal waves.
We assumed thermal inertia of the ground ice to be constant 
and equal to 2000~J~m$^{-2}$~s$^{-1/2}$~K$^{-1}$.

We considered (in GCM simulations) the internal heat flux and/or 
the thermal heat flux F$_{\text{ground}}$  
conducted from an hypothetical underlying ocean to be zero\footnote{This is a reliable assumption for thick enough 
(typically $>$ 100~m) water ice covers expected on hard Snowball Earth-like planets.}.

\subsection{Radiative Transfer}

The GCM includes a generalized radiative 
transfer for a variable gaseous atmospheric composition made of a mix of CO$_{2}$, N$_2$ 
and H$_{2}$O (HITRAN 2012 database \citep{Roth:13}) using the correlated-k method \citep{Fu:92,Eyme:16}) suited for fast calculation.
For this, we decomposed atmospheric temperatures, pressures, and water vapor mixing ratio 
into the following respective 7~x~8~x~8 grid, similar to \cite{Char:13}: 
Temperatures~=~$\{$100,150,~..~,350,400$\}$~K ; 
Pressures~=~$\{$10$^{-6}$,10$^{-5}$,~..~,1,10$\}$~bars ;
H$_2$O Mixing Ratio~=~$\{$10$^{-7}$,10$^{-6}$,~..~,10$^{-2}$,10$^{-1}\}$
~mol of H$_2$O $/$ mol of air (H$_2$O+CO$_2$ here).

CO$_2$ collision-induced absorptions (CIA) and dimer absorption \citep{Word:10co2} ) were 
included in our calculations as in \cite{Forg:13} and \cite{Word:13}, 
as well as N$_2$-N$_2$ collision-induced absorption \citep{Rich:12} and its role on the pressure broadening. 
We also added H$_2$O self and foreign continuums calculated with the CKD model \citep{Clou:89}, with H$_2$O lines truncated at 25~cm$^{-1}$.

For the computation, we used 32 spectral bands in the thermal infrared and 36 in the visible domain.
16 non-regularly spaced grid points were used for the g-space integration, where g is the cumulated
distribution function of the absorption data for each band.
We used a two-stream scheme \citep{Toon:89} to take into account radiative effects 
of aerosols (CO$_2$ ice and H$_2$O clouds) and Rayleigh scattering (mostly by N$_2$ and CO$_2$ molecules), 
using the method of \cite{Hans:74}.

We note that, for the calculation of absorption coefficients, 
we used only main isotopes: $^{12}$C$^{16}$O$_2$ and $^1$H$_2$$^{16}$O and $^{12}$N$_2$. It 
has been shown in similar conditions \citep{Hale:09} that 
the radiative effect of isotopic composition should be small. 
Nonetheless, we take this opportunity to encourage dedicated studies 
about the influence of isotopic composition 
on the radiative properties of (thick) atmospheres. 

\subsection{H$_2$O and CO$_2$ physical properties}
\label{ice_properties}

Both CO$_2$ and H$_2$O cycles are included in the GCM used in this work. 

\begin{table}
\begin{center}
\begin{tabular}{ll}
   \hline
              &              \\
   Physical parameters & Values \\
              &              \\
   \hline
              &              \\
   H$_2$O ice Albedo  & 0.6 \\
   H$_2$O ice thermal conductivity  & 2.5~W$~$m$^{-1}~$K$^{-1}$ \\
   H$_2$O ice emissivity  & 1.0 \\
   Ground H$_2$O ice thermal inertia & 2000 J~m$^{-2}$~s$^{-1/2}$~K$^{-1}$ \\  
              &              \\
   \hline
              &              \\
   CO$_2$ ice Albedo  & 0.6 \\  
   CO$_2$ ice thermal conductivity  & 0.5~W$~$m$^{-1}~$K$^{-1}$ \\
   CO$_2$ ice emissivity  & 0.9 \\
   CO$_2$ ice density  & 1.5 \\
              &              \\
   \hline
              &              \\
   Surface Topography & Flat \\
   Surface roughness coefficient & 0.01 m \\
              &              \\
   \hline

\end{tabular}
\end{center}
\caption{Physical Parameterizations used for the GCM calculations.}
\label{param_ices}
\end{table}

1. Melting, freezing, condensation, evaporation, sublimation and 
precipitation of H$_2$O physical processes are all included in the model.
In the atmosphere, water vapor can condense into water ice particles clouds.
At the surface, we fix the H$_2$O ice albedo at 0.6\footnote{This is the standard value used in the Snowball
Model Intercomparison (SNOWMIP) project \citep{Pier:11areps}.} and we use an emissivity of 1.

2. In our model, CO$_2$ can condense to form CO$_2$ ice clouds and surface frost if the temperature 
drops below the saturation temperature. Atmospheric CO$_2$ ice particles are sedimented 
and thus can accumulate at the surface. The CO$_2$ ice layer 
formed at the surface can sublimate and recycle the CO$_2$ in the atmosphere.
The CO$_2$ ice on the surface contributes to the surface albedo calculation: 
if the CO$_2$ ice layer exceeds a threshold value of 1~mm thickness, 
then the local surface albedo is set immediately to the albedo of CO$_2$ ice (0.6 in this work). 
On Mars, the albedo of CO$_2$ ice can substantially vary \citep{Kief:00jgr} with insolation and presence of dust. 
Without dust, the albedo can become very high so that 0.6 is probably a lower estimate.
For CO$_2$ ice, we use an emissivity of 0.9. 
The radiative effect of CO$_2$ ice clouds is discussed in details in section~\ref{co2_cloud_section}

Physical parameters used for both CO$_2$ and H$_2$O ices are summarized in table~\ref{param_ices}.

\section{Computation of maximal CO$_2$ ice thickness before basal melting} 
\label{max_ice_basal_melting}

Compared to water ice, the temperature dependence of CO$_2$ ice thermal conductivity is rather flat in the 100-250~K range (\citealt{Schm:97}; Part I, Figure~4).
We can thus estimate with a good approximation the maximum CO$_2$ ice thickness $h_{\text{max}}$ (limited by basal melting) by: 
\begin{equation}
 h_{\text{max}}~=~\frac{\lambda_{\text{CO}_2}~(T_{\text{melt}}-T_{\text{surf}})}{F_{\text{geo}}},
\label{conductivity_appendix}
\end{equation}
where $\lambda_{\text{CO}_2}$ is the thermal conductivity of CO$_2$ ice, $T_{\text{surf}}$ is the temperature at the top of the glacier and 
$T_{\text{melt}}$ is the melting temperature of CO$_2$ ice at the base of the glacier.

We derive then $T_{\text{melt}}$ using the Clausius-Clapeyron relation (CO$_2$ solid-liquid equilibrium) at the base of the glacier: 
\begin{equation}
 T_{\text{melt}}~=~T_{\text{ref}}~e^{\frac{(\frac{1}{\rho_{\text{liq}}}-\frac{1}{\rho_{\text{sol}}})}
{L_{\text{fus}}}~(g \rho_{\text{sol}} h_{\text{max}}+P_{\text{CO}_2}+P_{\text{N}_2}-P_{\text{ref}})},
\label{clapeyron_fus_appendix}
\end{equation}
with $\rho_{\text{sol}}$ and $\rho_{\text{liq}}$ the volumetric mass of liquid and solid CO$_2$, $L_{\text{fus}}$ the latent heat of 
fusion of CO$_2$ ice, $P_{\text{ref}}$ and $T_{\text{ref}}$ the coordinates of the triple point of CO$_2$, 
and $P_{\text{CO}_2}$ and $P_{\text{N}_2}$ the partial surface pressures of CO$_2$ and N$_2$, respectively. 
The pressure at the base of the glacier $P_{\text{melt}}$ is estimated 
from the equation $h_{\text{max}}$~=~$\frac{(P_{\text{melt}}-P_{\text{surf}})}{g~\rho_{\text{sol}}}$, 
with $P_{\text{surf}}$~=~$P_{\text{CO}_2}+P_{\text{N}_2}$.
We remind that we choose pN$_2$~=~1~bar for all our simulations.

We now derive the surface temperature $T_{\text{surf}}$ as a function of the surface pressure $P_{\text{surf}}$, 
using another Clausius-Clapeyron relation (CO$_2$ solid-gas equilibrium):
\begin{equation}
 T_{\text{surf}}~=~\frac{1}{\frac{1}{T_{\text{ref}}}-\frac{R}{L_{\text{sub}}~M_{\text{CO}_2}}~\ln(\frac{P_{\text{CO}_2}}{P_{\text{ref}}})},
\label{clapeyron_sub_appendix}
\end{equation}
with $L_{\text{sub}}$ the latent heat of sublimation of CO$_2$ ice and $M_{\text{CO}_2}$ the molar mass of CO$_2$.
It is in fact assumed here that the global mean temperature 
at the top of the CO$_2$ polar ice caps is constant and equal to the temperature of condensation of CO$_2$. This approximation should remain valid 
as soon as there is enough CO$_2$ in the atmosphere (or CO$_2$ stays a dominant atmospheric species).

From the set of equations~\ref{conductivity_appendix}-\ref{clapeyron_sub_appendix}, 
we get a new equation on $h_{\text{max}}$ of the form $h_{\text{max}}~=~e^{h_{\text{max}}}$, 
after several variable changes. We solve explicitely this equation using the lambert W function, and we obtain the following 
expression of the maximum possible CO$_2$ ice cap thickness $h_{\text{max}}$: 
\begin{equation}
\begin{cases}
 h_{\text{max}}~=~\frac{\lambda_{\text{CO}_2}}{F_{\text{geo}}}~\Big(\frac{1}{\frac{1}{T_{\text{ref}}}-\frac{R}{L_{\text{sub}}~M_{\text{CO}_2}}~\ln(\frac{P_{\text{surf}}}{P_{\text{ref}}})}\Big)\\
-\frac{L_{\text{fus}}}{(\frac{1}{\rho_{\text{liq}}}-\frac{1}{\rho_{\text{sol}}})~g \rho_{\text{CO}_2}}
\times~W\Big( \frac{-(\frac{1}{\rho_{\text{liq}}}-\frac{1}{\rho_{\text{sol}}})~g \rho_{\text{CO}_2} 
~\lambda_{\text{CO}_2}~T_{\text{ref}}~e^{\frac{(\frac{1}{\rho_{\text{liq}}}-\frac{1}{\rho_{\text{sol}}})}{L_{\text{fus}}}~(P_{\text{surf}}-P_{\text{ref}})}}
{L_{\text{fus}}~F_{\text{geo}}}\\
\times~e^{\frac{(\frac{1}{\rho_{\text{liq}}}-\frac{1}{\rho_{\text{sol}}})~g \rho_{\text{CO}_2}
\frac{\lambda_{\text{CO}_2}}{F_{\text{geo}}}}{L_{\text{fus}}}~\Big(\frac{1}{\frac{1}{T_{\text{ref}}}-\frac{R}{L_{\text{sub}}~M_{\text{CO}_2}}~\ln(\frac{P_{\text{surf}}}{P_{\text{ref}}})}\Big)}
\Big)\\
\end{cases}
\label{co2_ice_thickness_calc_appendix}
\end{equation}
This equation is used to directly compute $h_{\text{max}}$ as 
a function of the internal heat flux and CO$_2$ partial pressure (see Figure~\ref{2D_basal_melting}).

Note that we assumed a CO$_2$ ice volumetric mass density of 1500~kg~m$^{-3}$ 
and a CO$_2$ ice thermal conductivity of 0.5~W$~$m$^{-1}~$K$^{-1}$ 
(\citealt{Schm:97}; Part I, Figure~4).

\section{Stability of CO$_2$ ice caps}
\label{stability_co2}

What happens once CO$_2$ ice caps have reached their maximum size, as calculated in section~\ref{max_reservoir_co2} ?
Using calculations, we show in this appendix that, depending on a few parameters, the CO$_2$ ice caps when full may (or may not) 
be stable.

The total mass of CO$_2$ available at the surface and in the atmosphere is:
\begin{equation}
M_{\text{tot}} = M_{\text{atm}}+M_{\text{surf}} = \frac{S P_{\text{CO}_2}}{g}+A h_{\text{mean}} \rho_{\text{CO}_2},
\label{cap_stability1}
\end{equation}
with $A$ the area of the CO$_2$ ice caps, S the total area of the surface (i.e. 4$\pi$R$_\text{p}^2$) and 
$h_{\text{mean}}$ the thickness of the CO$_2$ ice caps averaged over $A$. 
When CO$_2$ ice caps are full, $h_{\text{mean}}~=~h_{\text{mean}}^{{\text{max}}}$.

We assume now that the system -- initially with full CO$_2$ ice caps -- is slightly perturbated from $P_{\text{CO}_2}~=~P$ to $P+\delta P$. 
By conservation of CO$_2$ mass, the new mass of CO$_2$ ice caps is 
$M_{\text{surf}}[P+\delta P]$~=~$\rho_{\text{CO}_2}~A h_{\text{mean}}[P+\delta P]\footnote{Hereafter, 
'[' and ']' are used to bracket variables.} $. 

If this quantity is lower than the maximum mass of CO$_2$ ice caps, 
e.g. $\rho_{\text{CO}_2}~A h_{\text{mean}}^{{\text{max}}}[P+\delta P]$, then the system is stable. 
Otherwise, the condition of unstability can be written formally:

\begin{equation}
\rho_{\text{CO}_2}~A h_{\text{mean}}[P+\delta P]>\rho_{\text{CO}_2}~A h_{\text{mean}}^{{\text{max}}}[P+\delta P],
\label{cap_stability2}
\end{equation}

which can be rewritten, using mass conservation:

\begin{equation}
M_{\text{tot}}-\frac{S ~ [P+\delta P]}{g}
>\rho_{\text{CO}_2}~A h_{\text{mean}}^{{\text{max}}}[P+\delta P], 
\label{cap_stability3}
\end{equation}

and using Taylor series: 

\begin{equation}
M_{\text{tot}}-\frac{S ~ P}{g}-\frac{S ~ \delta P}{g}
>\rho_{\text{CO}_2}~A h_{\text{mean}}^{{\text{max}}}[P]+\delta P~\rho_{\text{CO}_2}~\frac{d ( A h_{\text{mean}}^{{\text{max}}})}{dP_{\text{CO}_2}}[P].
\label{cap_stability4}
\end{equation}

Assuming that CO$_2$ ice caps were initially full, this yields to $h_{\text{mean}}[P]~=~h_{\text{mean}}^{{\text{max}}}[P]$ and gives:

\begin{equation}
\Big(M_{\text{tot}}-\frac{S ~ P}{g}-\rho_{\text{CO}_2}~A h_{\text{mean}}[P]\Big)-\frac{S ~ \delta P}{g}
>\delta P~\rho_{\text{CO}_2}~\frac{d ( A h_{\text{mean}}^{{\text{max}}})}{dP_{\text{CO}_2}}[P].
\label{cap_stability5}
\end{equation}

For $\delta P>0$, we have then:

\begin{equation}
A~\frac{d h_{\text{mean}}^{{\text{max}}}}{dP_{\text{CO}_2}}[P]+ h_{\text{mean}}^{{\text{max}}}~\frac{dA}{dP_{\text{CO}_2}}[P]<-\frac{S}{\rho_{\text{CO}_2}~g}.
\label{cap_stability6}
\end{equation}

We know first from GCM simulations that the area of the ice caps $A$ is not a monotonic function of $P_{\text{CO}_2}$ (see Figure~\ref{condens_plots}).
$A$ is in fact extremely rich in information, because it 
depends on a subtle combination of the greenhouse effect of CO$_2$, the surface thermal emission, 
the condensation temperature of CO$_2$ and the global heat atmospheric redistribution. 
As $P_{\text{CO}_2}$ grows, GCM simulations tell us that $\frac{dA}{dP_{\text{CO}_2}}$ should vary 
from positive to negative values. 
Fortunately, we also know from the various GCM simulations shown in Figure~\ref{condens_plots}b (23.5$^{\circ}$ obliquity, radiatively inactive CO$_2$ clouds) 
that the maximum of $A$($P_{\text{CO}_2}$) should for example roughly lie at a CO$_2$ partial pressure of 1~bar in this specific configuration.

It is possible to derive an analytical - complicated yet - expression of $\frac{dh_{\text{mean}}^{{\text{max}}}}{dP_{\text{CO}_2}}$ using 
equations~\ref{co2_ice_thickness_calc_appendix} and \ref{solution_glacier2}.
It tells us that, whatever the configuration, $\frac{dh_{\text{mean}}^{{\text{max}}}}{dP_{\text{CO}_2}}$ is always negative. 
As pCO$_2$ increases, the temperature at the top of the CO$_2$ ice caps also increases, 
which limits the thickness of the CO$_2$ ice caps\footnote{Equation~\ref{solution_glacier2} shows that the mean 
thickness of the CO$_2$ ice caps is proportional to the maximum thickness calculated by basal melting.} 
(see equation~\ref{conductivity}).

\begin{figure}
\begin{center}
 \includegraphics[width=12cm]{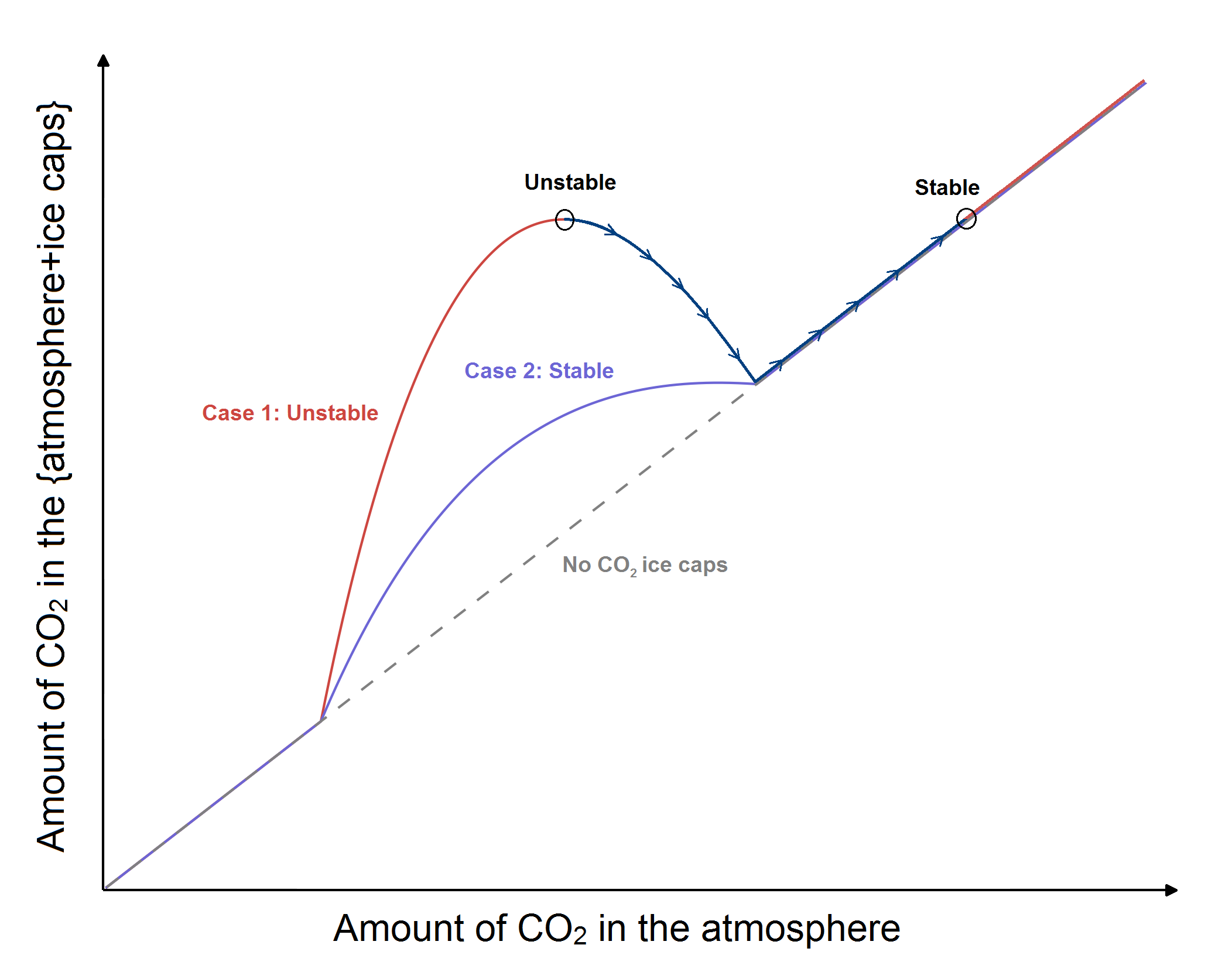}
\caption{Qualitative estimates of the maximum amount of CO$_2$ in the $\{$surface, atmosphere$\}$ as 
a function of the CO$_2$ partial pressure, for two different scenarios. In the first scenario (case 1), the CO$_2$ ice caps when full 
would be unstable. All the CO$_2$ ice get sublimed, forming a dense CO$_2$ atmosphere. In the second scenario (case 2), 
the CO$_2$ ice caps when full should progressively shrink as the CO$_2$ partial pressure increases, but would not be unstable. This 
configuration occurs when the decrease of the size of the CO$_2$ ice caps is offset by the increase of the CO$_2$ atmospheric mass.}
\label{ice_cap_stability}
\end{center}
\end{figure}

We illustrate now the previous calculations with the experiment extensively described in this work (1.30~AU, 23.5$^{\circ}$ obliquity, 
radiatively inactive CO$_2$ ice clouds, CO$_2$ partial pressure of 1~bar). Based on GCM simulations, we assume that 
$\frac{dA}{dP_{\text{CO}_2}}|_{pCO_2=1bar}~\sim~0$~m$^2$~Pa$^{-1}$ and 
we have therefore (with $\frac{A}{S}~\sim~0.1$; see section~\ref{co2_rad_extent}): 
\begin{equation}
\frac{dh_{\text{mean}}^{{\text{max}}}}{dP_{\text{CO}_2}} < -\frac{S}{g A \rho_{\text{CO}_2}} = \frac{-10}{9.81\times1.5\times10^3}=-7\times10^{-4}~\text{m}~\text{Pa}^{-1}.
\label{cap_stability7}
\end{equation}
Our calculations indicate that this condition is valid for an internal 
heat flux roughly lower than 80~mW~m$^{-2}$ (case 1 in Figure~\ref{ice_cap_stability}). 
It means that, for an internal heat flux lower than 80~mW~m$^{-2}$: (1) CO$_2$ ice caps when full would not be stable, and 
(2) the maximum amount of CO$_2$ that can be trapped in the $\{$surface, atmosphere$\}$ system 
is attained for pCO$_2$ lower or equal to 1~bar.

Conversely, for an internal heat flux higher than 80~mW~m$^{-2}$ (case 2 in Figure~\ref{ice_cap_stability}), 
CO$_2$ ice caps when full would be stable (unless $\frac{dA}{dP_{\text{CO}_2}}$ term becomes somehow significant).

\end{document}